\documentclass[a4paper, 11pt]{article}
\pdfoutput=1

\usepackage{amsmath}
\usepackage{amssymb}
\usepackage{amsfonts}
\usepackage{subfigure}
\usepackage[a4paper]{geometry}
\usepackage[labelfont=it, font=small, labelsep=colon]{caption}
\usepackage{graphicx}
\usepackage[usenames,dvipsnames]{xcolor}
\usepackage{natbib}
\usepackage[breaklinks, colorlinks, linkcolor=black, citecolor=black,
urlcolor=black]{hyperref}
\usepackage{hypernat}

\def\Keywords{Sparse Regression, Feature Selection, Iterative Majorization}

\hypersetup{%
	pdfauthor={Gerrit J.J. van den Burg, Patrick J.F. Groenen, Andreas 
		Alfons},
	pdftitle={SparseStep: Approximating the Counting Norm for Sparse 
		Regularization},
	pdfkeywords={\Keywords},
	pdfsubject={Statistical Methodology}
}

\usepackage{algorithm}
\usepackage{algorithmic}

\bibpunct{(}{)}{,}{a}{,}{,}

\graphicspath{{./images/}}

\newcommand{\bft}[1]{\mathbf{#1}}
\newcommand{\bfs}[1]{\boldsymbol{#1}}
\newcommand{\abs}[1]{\left|#1\right|}
\DeclareMathOperator*{\argmin}{arg\,min}

\title{SparseStep: Approximating the Counting Norm for Sparse Regularization}
\author{
	Gerrit J.J. van den Burg \\ burg@ese.eur.nl
	\and
	Patrick J.F. Groenen \\ groenen@ese.eur.nl
	\and
	Andreas Alfons \\ alfons@ese.eur.nl
}
\date{Econometric Institute, Erasmus University Rotterdam, The 
	Netherlands\\[2ex]\today}

\begin{document}
\maketitle

\begin{abstract}
	\noindent
	The SparseStep algorithm is presented for the estimation of a sparse
	parameter vector in the linear regression problem. The algorithm works
	by adding an approximation of the exact counting norm as a constraint
	on the model parameters and iteratively strengthening this
	approximation to arrive at a sparse solution. Theoretical analysis of 
	the penalty function shows that the estimator yields unbiased 
	estimates of the parameter vector.  An iterative majorization 
	algorithm is derived which has a straightforward implementation 
	reminiscent of ridge regression. In addition, the SparseStep algorithm 
	is compared with similar methods through a rigorous simulation study 
	which shows it often outperforms existing methods in both model fit 
	and prediction accuracy.
	\vskip\baselineskip\noindent
	\textbf{Keywords:} \Keywords
\end{abstract}

\section{Introduction}
\label{sec:intro}
In many modeling problems it is desirable to restrict the number of nonzero
elements in the parameter vector to reduce the model complexity. Achieving
this so-called sparsity in the model parameters for the regression problem is
known to be NP-hard \citep{natarajan1995sparse}. Many alternatives have been
presented in the literature which approximate the true sparse solution by
applying shrinkage to the parameter vector, such as for instance the lasso
estimator \citep{tibshirani1996regression}.  However, this shrinkage can
underestimate the true effect of explanatory variables on the model outcome.
Here, an algorithm is presented which creates sparse model estimates but does
not apply shrinkage to the parameter estimates. This feature is obtained by 
approximating the exact counting norm and making this approximation 
increasingly more accurate.

Traditional methods for solving the exact sparse linear regression problem
include best subset selection, forward and backward stepwise regression, and
forward stagewise regression. These approaches may not always be feasible for 
problems with a large number of predictors and may display a high degree of 
variance with out-of-sample predictions \citep{hastie2009elements}.  
Alternatively, penalized least-squares methods add a regularization term to 
the regression problem, to curb variability through shrinkage or induce 
sparsity, or both. Of the many more recent approaches to this problem, the 
most well-known are perhaps the SCAD penalty \citep{fan2001variable} and the 
MC+ penalty \citep{zhang2010nearly}.  In both of these approaches, a
penalty is added such that the overall size of the model parameters can be 
controlled.

\begin{figure*}[tb]
	\centering
	\def\FigureScale{0.66}
	\subfigure[Ridge \& Lasso]{%
		\includegraphics[scale=\FigureScale]{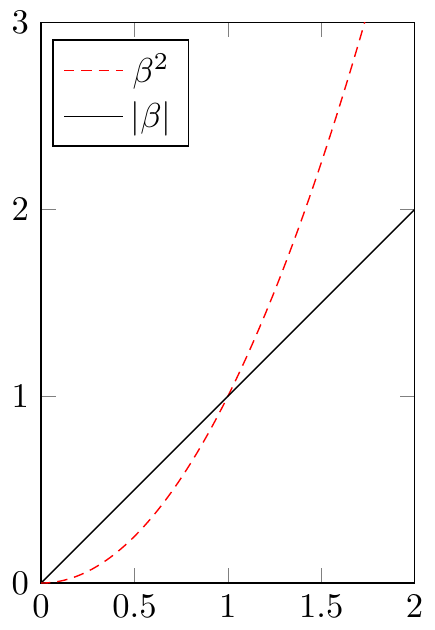}
		\label{fig:comp_classic}
	}\qquad
	\subfigure[SCAD]{%
		\includegraphics[scale=\FigureScale]{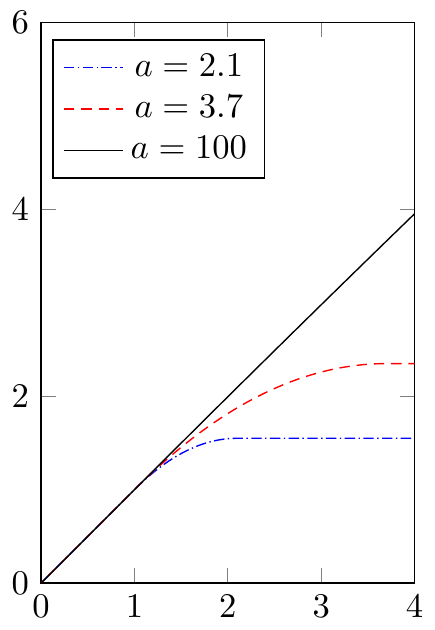}
		\label{fig:comp_scad}
	}\qquad
	\subfigure[MC+]{%
		\includegraphics[scale=\FigureScale]{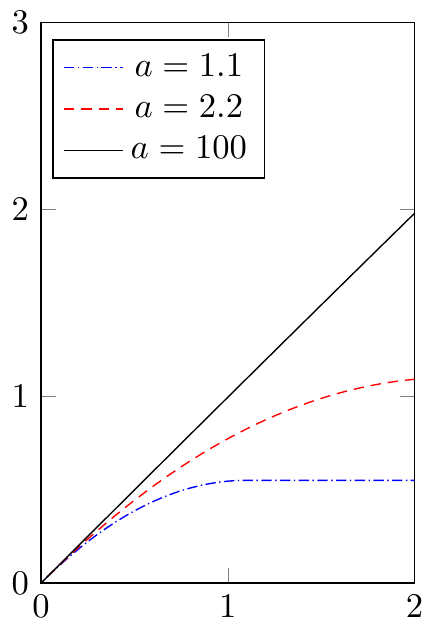}
		\label{fig:comp_mcplus}
	}\qquad
	\subfigure[SparseStep]{%
		\includegraphics[scale=\FigureScale]{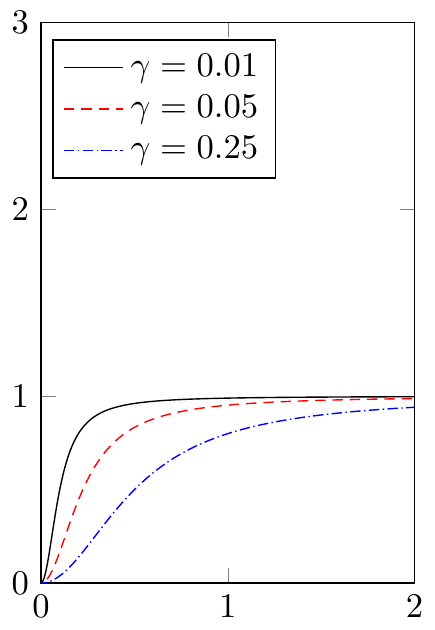}
		\label{fig:comp_sparsestep}
	}
	\caption{Illustrations of different penalty functions, with 
		regularization parameter $\lambda = 1$ where applicable.  With 
		the exception of the Ridge penalty all penalties can induce 
		sparsity in the parameter vector. All penalties are symmetric 
		around the origin. Note that out of all penalties shown here 
		only the Ridge and SparseStep penalties have continuous first 
		derivatives at the origin. For the SCAD penalty the axes are 
		rescaled to accommodate the requirement that $a > 2$ for this 
		penalty.  \label{fig:comparison}}
\end{figure*}

Figure \ref{fig:comparison} shows an illustration of the different penalty 
functions discussed above. Note that all penalty functions are symmetric 
around zero, including the SparseStep penalty introduced below. It can be seen 
that the shapes of the SCAD and MC+ penalties closely resemble each other. The 
different shapes of the penalty function for the SCAD and MC+ penalty are due 
to the parameter $a$, which can be optimized over for a given dataset. In 
contrast, the different shapes of the SparseStep penalty show subsequent 
approximations of the exact $\ell_0$ penalty, as described below.

This paper is organized as follows. In Section \ref{sec:theory} the theory 
behind the SparseStep penalty is introduced and analyzed.  Section 
\ref{sec:methods} derives the SparseStep algorithm using the Iterative 
Majorization technique and describes the implementation of the algorithm.  
Experiments comparing SparseStep with existing methods are described 
extensively in the following section. Section \ref{sec:discussion} concludes.

\section{Theory}
\label{sec:theory}
Below the theory of norms and regularized regression is briefly reviewed,
after which the SparseStep norm approximation is presented and analyzed.

\subsection{Norms}
\begin{figure}[tb]
	\centering
	\includegraphics[scale=1.5]{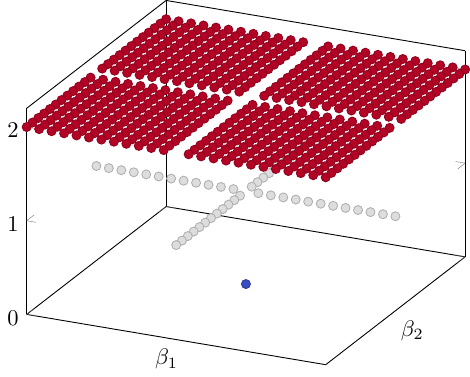}
	\caption{Illustration of the exact $\ell_0$ counting norm in two
		dimensions. Note the strong discontinuities along both axes, 
		as indicated by the color of the points.  The variables 
		$\beta_1$ and $\beta_2$ may vary continuously along the axes, 
		but for convenience the counting norm is computed here on a 
		grid of values. \label{fig:counting}}
\end{figure}

Let the $\ell_p$ norm of a vector $\bfs{\beta} \in \mathbb{R}^m$ be defined as
\[
	\| \bfs{\beta} \|_p \equiv \left( \sum_{j=1}^m |\beta_j|^p
	\right)^{1/p}.
\]
For $p = 2$, the well known Euclidean norm is obtained, whereas for $p = 1$
the distance measured is known as the Manhattan distance. When $p = 0$, using 
the definition $0^0 = 0$, this function seizes to be a proper norm due to the 
lack of homogeneity and is equal to the number of nonzero elements of 
$\bfs{\beta}$ \citep{peetre1972interpolation, donoho2006compressed}.  
Therefore, let us denote by $\ell_0$ the \emph{pseudonorm} given by,
\[
	\| \bfs{\beta} \|_0 = \sum_{j=1}^m \pi[\beta_j \neq 0],
\]
where $\pi(\cdot)$ is an indicator function which is 1 if it's argument is
true and 0 otherwise. For simplicity the $\ell_0$ pseudonorm will be referred 
to as the counting \emph{norm} throughout this paper, even though it is not a 
proper norm in the mathematical sense. The counting norm is shown graphically 
in Figure \ref{fig:counting} for the two dimensional case.  It can be seen 
that this norm is discontinuous and nonconvex.

\subsection{Regularization}
Let $\mathcal{D} = \{ (\bft{x}_i', y_i)\}_{i=1,\ldots,n}$ denote the data for 
the regression problem, with explanatory variables $\bft{x}_i = (x_{i1}, 
\ldots, x_{im})' \in \mathbb{R}^m$ and outcome $y_i \in \mathbb{R}$. Let 
$\bft{X} \in \mathbb{R}^{n\times m}$ denote the data matrix with rows 
$\bft{x}_i'$.  Assume that the vector of outcomes $\bft{y} \in \mathbb{R}^n$ 
is centered, so that the intercept term can be ignored.  The least-squares 
regression problem can then be written as
\[
	\hat{\bfs{\beta}} = \argmin_{\bfs{\beta}} \| \bft{y} - 
	\bft{X}\bfs{\beta} \|^2.
\]
The $\ell_p$-\emph{regularized} least-squares problem can be defined through 
the loss function
\[
	L(\bfs{\beta}) = \| \bft{y} - \bft{X}\bfs{\beta} \|^2 + \lambda 
	\sum_{j=1}^m |\beta_j|^p,
\]
with $\lambda \geq 0$ a regularization parameter. Two well-known special cases 
of these regularized least-squares problems are ridge regression 
\citep{hoerl1970ridge} corresponding to $p=2$ and the lasso estimator 
\citep{tibshirani1996regression} corresponding to $p=1$. When $p = 0$ is used 
the regularization term turns into the counting norm of the $\beta_j$. Note 
that in this case no shrinkage of the $\beta_j$ occurs because the number of 
nonzero $\beta_j$ is controlled and not their size.

\subsection{Norm Approximation}
\begin{figure*}[tb]
	\centering
	\subfigure[$\gamma^2 = 0.3$]{%
		\includegraphics[scale=0.95]{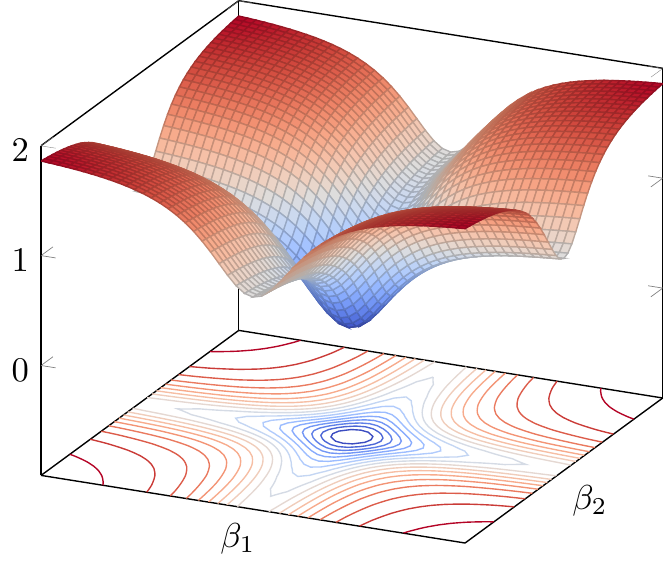}
		\label{fig:approx2l}
	}\qquad
	\subfigure[$\gamma^2 = 0.05$]{%
		\includegraphics[scale=0.95]{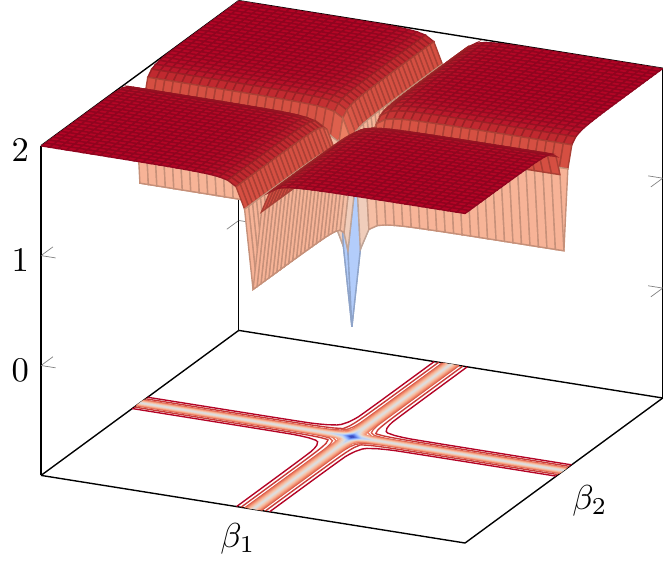}
		\label{fig:approx2s}
	}
	\caption{Three-dimensional illustrations of the norm approximation
		proposed by \citet{de2011deconvolution} for two different
		values of $\gamma$. It can be seen that for a smaller value of
		$\gamma$ the approximation more closely resembles the exact
		$\ell_0$ norm illustrated in Figure \ref{fig:counting}.  
		\label{fig:approx}}
\end{figure*}

Recently, an approximation to the $\ell_0$ norm was proposed by
\citet{de2011deconvolution}, where the indicator function of an element
$\beta_j$ is approximated as
\[
	|\beta_j|^0 = \pi[\beta_j \neq 0 ] \approx \frac{\beta_j^2}{\beta_j^2 
		+ \gamma^2},
\]
where $\gamma \geq 0$ is a positive constant\footnote{For consistency with the
	remainder of the paper our definition of $\gamma$ deviates from that
	of \citet{de2011deconvolution}. In contrast to their definition of the
	approximation, the square of $\gamma$ is used here.}.  Note that if
$\gamma = 0$ the approximation becomes exact. By decreasing the value of the 
$\gamma$ parameter the approximation of the counting norm becomes increasingly 
more accurate. Figures~\ref{fig:approx2l} and \ref{fig:approx2s} show the 
approximation for both large and small values of $\gamma$, respectively. It
can be seen that for decreasing values of $\gamma$ the approximation indeed
converges to the exact counting norm shown in Figure~\ref{fig:counting}.

In the following, the approximation will be used to define the 
\emph{SparseStep penalty} as
\[
	P_{\lambda}(\beta_j) = \lambda \frac{\beta_j^2}{\beta_j^2 + \gamma^2}.
\]
To prove that this penalty results in unbiased estimates of the true $\beta_j$
parameter, the approach of \citet{fan2001variable} is followed. For 
unbiasedness it is required that the derivative of the penalty term is zero 
for large values of $\abs{\beta_j}$. The derivative of the SparseStep penalty 
is
\[
	P_{\lambda}'(\beta_j) = \lambda \frac{2\gamma^2\beta_j}{( \beta_j^2 + 
		\gamma^2)^2},
\]
hence, it must hold that
\[
	\lim_{\abs{\beta_j} \rightarrow \infty} P_{\lambda}'(\abs{\beta_j}) =
	\lim_{\abs{\beta_j} \rightarrow \infty} \lambda 
	\frac{2\gamma^2\abs{\beta_j}}{(\abs{\beta_j}^2 + \gamma^2)^2} = 0,
\]
which is indeed the case, proving that the SparseStep penalty results in 
unbiased estimates.

Additionally, \citet{fan2001variable} derive sufficient conditions for a 
penalty function to have the Oracle Property. This property means that
under certain regularity conditions a method correctly identifies the sparsity 
in the predictor variables correctly, as the number of observations goes to 
infinity.  One sufficient condition for this is that the derivative of the 
penalty function should be positive at the origin.  This does not hold exactly 
for SparseStep, but does hold in an arbitrarily small region around the 
origin, due to the value of $\gamma$. We therefore conjecture that the Oracle 
Property also holds for SparseStep, but further research is necessary.

The above leads naturally to the formulation of the SparseStep regression 
problem, with loss function
\[
	L(\bfs{\beta}) =  \| \bft{y} - \bft{X}\bfs{\beta} \|^2 + \lambda 
	\sum_{j=1}^m \frac{\beta_j^2}{\beta_j^2 + \gamma^2}.
\]
In the next section, an Iterative Majorization algorithm will be derived for 
minimizing this loss function.

\section{Methodology}
\label{sec:methods}
With the theoretical underpinnings of SparseStep established above, it is now
possible to derive the optimization algorithm necessary to minimize the
SparseStep regression loss function. The approach used here is that of the
\emph{iterative majorization} algorithm. A brief introduction is given first,
followed by the derivation of the SparseStep algorithm.

\subsection{Iterative Majorization}
The Iterative Majorization (IM) algorithm is a general optimization algorithm
based on surrogate functions, first described by \citet{ortega1970iterative}.
It is also known as the Majorization Minimization algorithm and is a 
generalization of the popular Expectation Maximization algorithm \citep[see 
e.g.][]{hunter2004tutorial}. A brief description of the algorithm follows.

Let $f : \mathcal{X} \rightarrow \mathcal{Y}$ with $\mathcal{X} \subseteq
\mathbb{R}^d$ and $\mathcal{Y} \subseteq \mathbb{R}$ be the function that
needs to be optimized. Construct a majorizing function $g : \mathcal{X}
\times \mathcal{X} \rightarrow \mathcal{Y}$ such that
\begin{align*}
	f(y) &= g(y, y), \\
	f(x) &\leq g(x, y) \textrm{ for all } x, y \in \mathcal{X},
\end{align*}
where $y$ is the so-called \emph{supporting point}. Differentiability of $f$
at $y$ implies that $\nabla f(y) = \nabla g(y, y)$. Given a majorizing
function $g$ the following procedure results in a stationary point of
$f$:
\begin{enumerate}
	\item Let $y = x_0$, with $x_0$ a starting point
	\item Minimize $g(x, y)$ and let $x^+ = \argmin g(x, y)$
	\item Stop if a stopping criterion is reached, otherwise let $y = x^+$ 
		and go to step 2.
\end{enumerate}
This procedure yields a guaranteed descent algorithm where $f(x^+) \leq f(y)$, 
with a linear convergence rate \citep{de1994block}. However, a well-known 
property of the IM algorithm is that in the first few iterations
often large improvements in the loss function can be made
\citep{havel1991evaluation}.  This property makes it ideally suited for the
SparseStep algorithm described below. Generally, a sufficiently simple
functional form is chosen for the majorizing function such that Step 2 in
the above procedure can be done swiftly. In the case of the SparseStep
regression problem, a quadratic majorizing function is most appropriate.

\subsection{Majorization Derivation}
\begin{figure}[tb]
	\centering
	\includegraphics[scale=0.95]{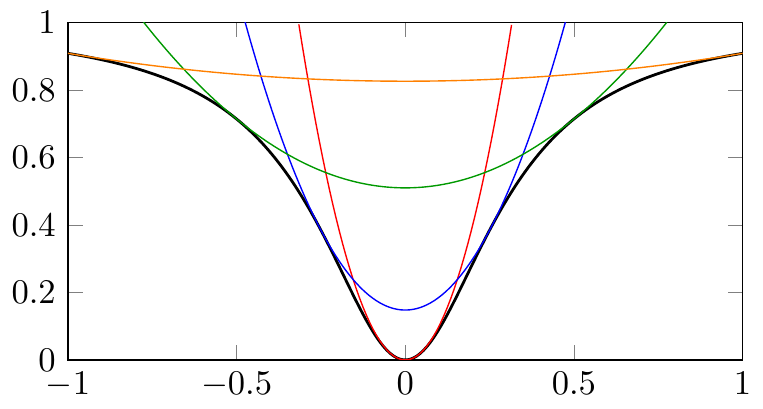}
	\caption{Illustrations of the majorizing function $g(x,y)$ 
		(interrupted lines) of the SparseStep penalty $f(x)$ (solid 
		line), with $\gamma^2 = 0.1$. The majorizing function is shown 
		for various values of the supporting point: the dotted red 
		line at $y = 0$, the dash dotted blue line at $y = 0.25$, and 
		the dashed green line at $y = 0.5$. The vertical lines mark 
		these supporting points for the latter two values.  
		\label{fig:major_illustrations}}
\end{figure}

The majorizing function of the SparseStep penalty function will be derived
here. For ease of notation let $f(x)$ denote the penalty function, with $x
\in \mathbb{R}$ and let $g(x, y)$ denote the majorizing function, with $x, y 
\in \mathbb{R}$. Then,
\begin{align*}
	f(x) &= \frac{x^2}{x^2 + \gamma^2}, \\
	g(x, y) &= ax^2 - 2b x + c,
\end{align*}
where the coefficients of $g(x, y)$ generally depend on $y$. Taking first
derivatives yields
\begin{align*}
	f'(x) &= \frac{2x\gamma^2}{(x^2 + \gamma^2)^2}, \\
	g'(x, y) &= 2ax - 2b.
\end{align*}
Since the SparseStep penalty is symmetric, it is desirable that $g(x, y)$ is 
symmetric as well and thus has its minimum at $x = 0$. From this it follows 
that $g'(0, y) = 0$, which implies $b = 0$. Next, the majorizing function must 
be tangent to the penalty function at the supporting point $x = y$, thus it is 
required that $g'(y, y) = f'(y)$, which yields
\begin{align*}
	a &= \frac{\gamma^2}{(y^2 + \gamma^2)^2}.
\end{align*}
Finally, the majorizing function must have the same function value at the
supporting point, thus $g(y, y) = f(y)$, this gives
\begin{align*}
	\frac{\gamma^2}{(y^2 + \gamma^2)^2}\, y^2 + c &= \frac{y^2}{y^2 +
		\gamma^2} \\
	c &= \frac{y^4}{(y^2 + \gamma^2)^2}.
\end{align*}
Thus, $g(x,y)$ becomes
\[
	g(x, y) = \frac{\gamma^2}{(y^2 + \gamma^2)^2}x^2 + \frac{y^4}{(y^2 +
		\gamma^2)^2} = \frac{\gamma^2 x^2 + y^4}{(y^2 + \gamma^2)^2}.
\]
It now remains to be shown that the majorizing function is everywhere above
the penalty function, $g(x, y) \geq f(x)$ for all $x \in \mathbb{R}$.  This
can be done as follows,
\begin{align*}
	g(x, y) &\geq f(x) &\Leftrightarrow \\
	\frac{\gamma^2 x^2 + y^4}{(y^2 + \gamma^2)^2} &\geq \frac{x^2}{x^2 +
		\gamma^2}  &\Leftrightarrow\\
	(\gamma^2 x^2 + y^4)(x^2 + \gamma^2) &\geq x^2 (y^2 + \gamma^2)^2
	&\Leftrightarrow\\
	x^4 + y^4 - 2x^2y^2 &\geq 0  &\Leftrightarrow\\
	(x^2 - y^2)^2 &\geq 0. & \square
\end{align*}
Figure~\ref{fig:major_illustrations} shows the majorizing function and the 
SparseStep penalty function for different values of the supporting point. Note 
that the majorizing function is indeed symmetric around 0, as desired.

The above derivation ensures that a majorizing function for the SparseStep
loss function can be derived.  Recall that the SparseStep loss function is 
given by
\[
	L(\bfs{\beta}) = \| \bft{y} - \bft{X}\bfs{\beta} \|^2 + \lambda
	\sum_{j=1}^m \frac{\beta_j^2}{\beta_j^2 + \gamma^2}.
\]
Let $\alpha_j$ denote the previous value of $\beta_j$ in the IM algorithm (the
supporting point). Then, using the majorizing function derived above it is
clear that the following inequality holds
\[
	L(\bfs{\beta}) \leq \| \bft{y} - \bft{X}\bfs{\beta} \|^2 + \lambda
	\sum_{j=1}^m \frac{\gamma^2 \beta_j^2 + \alpha_j^4}{(\alpha_j^2 +
		\gamma^2)^2} = G(\bfs{\beta}, \bfs{\alpha}),
\]
where $G(\bfs{\beta}, \bfs{\alpha})$ denotes the majorizing function of 
$L(\bfs{\beta})$.  Taking the derivative of $G(\bfs{\beta}, \bfs{\alpha})$ 
with respect to $\bfs{\beta}$ yields an explicit expression for the update of 
$\bfs{\beta}$ in the IM algorithm.  Before taking the derivative of the 
majorizing function however, let us define
\begin{align}
	\bfs{\Omega}_{jj} &= \frac{\gamma^2}{(\alpha_j^2 + \gamma^2)^2}
	\label{eq:omegadef} \\
	\delta_j &= \alpha_j^2/\gamma, \nonumber
\end{align}
such that $\bfs{\Omega}$ is an $m \times m$ diagonal matrix with elements
$\bfs{\Omega}_{jj}$ and $\bfs{\delta} \in \mathbb{R}^m$ a vector with elements 
$\delta_j$.  With these definitions, the regularization term
becomes
\[
\lambda \sum_{j=1}^m \frac{\gamma^2 \beta_j^2 + \alpha_j^4}{(\alpha_j^2 +
	\gamma^2)^2} = \lambda (\bfs{\beta}' \bfs{\Omega} \bfs{\beta} +
\bfs{\delta}' \bfs{\Omega} \bfs{\delta}).
 \]
By expanding the norm and using this form for the regularization term, it
is possible to write $G(\bfs{\beta}, \bfs{\alpha})$ as
\[
	G(\bfs{\beta}, \bfs{\alpha}) = \bft{y}'\bft{y} -
	2\bft{y}'\bft{X}\bfs{\beta} + \bfs{\beta}'(\bft{X}'\bft{X} + \lambda
	\bfs{\Omega})\bfs{\beta} + \lambda
	\bfs{\delta}'\bfs{\Omega}\bfs{\delta}.
\]
Taking the derivative to $\bfs{\beta}$ and setting this to zero, yields
\[
	-2\bft{X}'\bft{y} + 2(\bft{X}'\bft{X} +
	\lambda\bfs{\Omega})\bfs{\beta} = 0
\]
Thus, the update of the majorization algorithm is simply
\[
	\bfs{\beta} = \left( \bft{X}'\bft{X} + \lambda\bfs{\Omega}
	\right)^{-1} \bft{X}'\bft{y}.
\]
Since $\bfs{\Omega}$ is a diagonal matrix this expression is remarkably
similar to the solution of the ridge regression problem, in which
$\bfs{\Omega}$ is simply the identity matrix.

\subsection{SparseStep Algorithm}
With the derivation of the IM algorithm for minimizing the SparseStep
regression loss function, it is now possible to formulate the SparseStep
algorithm. To avoid local minima, the SparseStep penalty is introduced slowly
by starting with a large value of $\gamma$, so that the penalty is very smooth 
and behaves like a ridge penalty. Then, the $\gamma$ value is reduced so that 
the irregularity and nonconvexity is introduced slowly. The value of $\gamma$ 
is reduced until it is close to zero.  By slowly introducing the nonsmoothness 
in the penalty and taking only a few steps of the IM algorithm for each 
$\gamma$, the SparseStep algorithm aims to avoid local minima and tries to 
reach the global minimum of the regression problem with the counting norm 
penalty.  The pseudocode for SparseStep regression is given in Algorithm 
\ref{alg:sparsestep}.

The algorithm starts by initializing $\bfs{\beta}$ and $\gamma$ from given 
values $\bfs{\beta}_0$ and $\gamma_0$ respectively.  For each value of 
$\gamma$ the parameter estimates are updated $t_{max}$ times using the IM 
algorithm. Subsequently $\gamma$ is reduced by a factor $\gamma_{step}$. This 
process is continued until $\gamma$ reaches a provided stopping value 
$\gamma_{stop}$. In the end, sufficiently small elements of $\bfs{\beta}$ are 
set to absolute zero by comparing to a small constant $\epsilon$. This is done 
to avoid numerical precision errors and can be a method for enhancing the 
sparsity inducing properties of SparseStep. The value of $\epsilon$ and that 
of $\gamma_{stop}$ are related. Note that in an actual implementation the 
matrices $\bft{X}'\bft{X}$ and $\bft{X}'\bft{y}$ should be cached for 
computational efficiency.

\begin{algorithm}[tb]
	\caption{SparseStep Regression}
	\label{alg:sparsestep}
	\begin{algorithmic}
		\STATE {\bfseries Input:} $\bft{X}$, $\bft{y}$,
		$\bfs{\beta}_0$, $\gamma_0$, $\gamma_{stop}$, $\gamma_{step}$,
		$t_{max}$, $\epsilon$
		\STATE {\bfseries Output:} $\bfs{\beta}$
		\STATE $\bfs{\beta} \leftarrow \bfs{\beta}_0$
		\STATE $\gamma \leftarrow \gamma_0$
		\WHILE{$\gamma > \gamma_{stop}$}
		\FOR{$t = 1$ to $t_{max}$}
		\STATE Construct $\bfs{\Omega}$ according to
		(\ref{eq:omegadef})
		\STATE $\bfs{\beta} \leftarrow \left( \bft{X}'\bft{X} +
			\lambda\bfs{\Omega} \right)^{-1} \bft{X}'\bft{y}$
		\ENDFOR
		\STATE $\gamma \leftarrow \gamma / \gamma_{step}$
		\ENDWHILE
		\FOR{$j \in \{1, \ldots, m\}$}
		\IF{$|\beta_j| < \epsilon$}
		\STATE $\beta_j \leftarrow 0$
		\ENDIF
		\ENDFOR
	\end{algorithmic}
\end{algorithm}

\section{Experiments}
\label{sec:experiments}

To verify the performance of the SparseStep algorithm in correctly identifying
the nonzero predictor variables in a regression problem, a simulation study
was performed. The aim of this simulation study is to mimic as much as
possible a practical setting where a researcher is interested in both the
predictive accuracy of a regression model and the correct identification of
variables with nonzero coefficients. Moreover, this simulation study allows 
verification of the performance of the SparseStep algorithm for datasets with 
varying statistical properties such as the number of variables, the 
signal-to-noise ratio (SNR), the correlation between the variables, and the 
degree of sparsity in the true coefficient vector.

A second goal of this simulation study is to compare the performance of
the SparseStep algorithm with existing regression methods. These competing 
methods are: ordinary least-squares, lasso \citep{tibshirani1996regression}, 
ridge regression \citep{hoerl1970ridge}, SCAD \citep{fan2001variable}, and MC+ 
\citep{zhang2010nearly}. Thus, the focus here is on comparing SparseStep with 
other penalized regression methods, including some that induce sparsity 
through penalization.  In order to accurately evaluate the predictive accuracy 
of these methods and to find the best regularization parameter for each 
method, separate training and testing datasets will be used. The procedure is 
then to find for each method the regularization parameter which performs best 
on the training dataset, as measured using 10-fold cross-validation. Next, 
each method is trained one more time on the entire training dataset using this 
optimal regularization parameter and the obtained model is used to predict the 
test dataset.  Predictive accuracy is measured using the mean squared error 
(MSE) on the test dataset.

The accuracy of the estimated parameter vector ${\bfs{\hat{\beta}}}$ will be
evaluated on two measures: the mean squared error with respect to the true
$\bfs{\beta}$ and the \emph{sparsity hitrate}. The sparsity hitrate is
calculated simply as the sum of the correctly identified zero elements of
$\bfs{\beta}$ and the correctly identified nonzero elements, divided by the
total number of elements of $\bfs{\beta}$.

For the simulation study the following data generating process will be used.
Let $\bft{X} \in \mathbb{R}^{n\times m}$ denote the data matrix drawn from a
multivariate normal distribution with mean vector $\bfs{\mu} \in \mathbb{R}^m$
and correlation matrix $\bfs{\Sigma} \in \mathbb{R}^{m\times m}$, such that
the rows $\bft{x}_i' \sim \mathcal{N}(\bfs{\mu}, \bfs{\Sigma})$. The data 
matrix $\bft{X}$ was scaled such that each column had mean 0 and unit 
variance. In all simulated datasets $\bfs{\mu}$ is drawn from an 
$m$-dimensional standard uniform distribution. For the correlation matrix 
$\bfs{\Sigma}$ three different scenarios are used: uncorrelated, constantly 
correlated, and noise correlated.  In the uncorrelated case the $\bfs{\Sigma}$ 
matrix is simply the identity matrix, in the constantly correlated case all 
variables have a correlation of .5 with each other, whereas in the noise 
correlated case a correlation matrix is generated by adding realistic noise to 
the identity matrix using the method of
\citet{hardin2013method}\footnote{This corresponds to Algorithm 4 in the paper
	of Hardin et al., using the default parameters of $\varepsilon = 0.01$
	and $M = 2$ (in their notation).}. Next, a parameter vector
$\bfs{\beta} \in \mathbb{R}^m$ is drawn from a uniform distribution with
elements $\beta_j \in [-1, 1]$. The last $z$ elements of $\bfs{\beta}$ are set
to zero to simulate sparsity. Finally, to obtain realistic data with a known
signal-to-noise ratio, the simulated outcome variable $\bft{y}$ is calculated
as
\[
	\bft{y} = \bft{X}\bfs{\beta} + \bft{e},
\]
where $\bft{e} \in \mathbb{R}^n$ is a noise term which contains $n$ elements 
drawn from a univariate normal distribution with mean zero and standard 
deviation such that the SNR given by 
$\bfs{\beta}'\bft{X}'\bft{X}\bfs{\beta}/\bft{e}'\bft{e}$ is as desired.

\begin{table}[tb]
	\centering
	\caption{Overview of the values for different parameters in the data
		generating process of the simulation study. Datasets were
		generated for each possible combination of these values,
		resulting in 180 datasets. \label{tab:simulation_overview}}
	\begin{tabular}{lr}
		{\bfseries Parameter} & {\bfseries Values} \\
		\hline
		Variables ($m$) & $10, 50, 100, 500$ \\
		Sparsity ($\zeta \%$) & $0, 25, 50, 75, 95$ \\
		SNR & $0.1, 1.0, 10.0$ \\
		Correlation & uncorrelated, constant, noise \\
		\hline
	\end{tabular}
\end{table}

Table \ref{tab:simulation_overview} gives an overview of how the different
parameters of the data generating process were varied among datasets. Using
these parameters a total of 180 datasets were generated. For all datasets the
number of instances $n$ was set to 30,000, which was then split into 20,000
instances in the training dataset and 10,000 in the testing dataset. Note that
the degree of sparsity in the table is expressed as a percentage of the number
of variables $m$. In practice, the number of zeroes in $\bfs{\beta}$
corresponds to $z = \lfloor m \cdot \zeta / 100\rfloor$, where $\zeta$ is a
number taken from the second row of the table.

The simulation study was set up such that each method was trained on exactly
the same cross-validation sample when the same value of $\lambda$ was 
supplied. The grid of $\lambda$ parameters for the regularized methods came 
from a logarithmically spaced vector of 101 values between $2^{-15}$ and 
$2^{15}$.  For SparseStep, the input parameters were chosen as $\gamma_0 = 
10^6$, $\gamma_{stop} = 10^{-8}$, $\gamma_{step} = 2$, $t_{max} = 2$, 
$\epsilon = 10^{-7}$, and $\bfs{\beta}_0 = \bft{0}$ (see Algorithm 
\ref{alg:sparsestep}).  Default input parameters where chosen for the other 
methods where applicable.  The R language \citep{team2015r} was used for the 
SCAD and MC+ methods, with SCAD implemented through the ncvreg package 
\citep{breheny2011coordinate}, and MC+ through the SparseNet package 
\citep{mazumder2011sparsenet}.  The other methods were implemented in the 
Python language \citep{rossum1995python} using the scikit-learn package 
\citep{pedregosa2011scikit}. For the MC+ penalty the secondary regularization 
parameter $a$ was optimized for the training dataset using the CV 
implementation of the SparseNet package. For SCAD $a$ was set to 3.7 as per 
the default settings of ncvreg and \citet{fan2001variable}.

To determine statistically significant differences between the performance of
each of the  methods, recommendations on benchmarking machine learning methods
will be used as formulated by \citet{demvsar2006statistical}.  Specifically,
\emph{rank tests} will be applied to evaluate whether SparseStep outperforms
other methods significantly. For each dataset, fractional ranks are calculated
for each performance measure with a smaller rank indicating a better
performance. Methods are considered to have equal performance if the 
difference on a performance metric is smaller than $10^{-4}$. A Friedman rank 
test can be done on the calculated ranks to test for equal performance of the 
methods \citep{friedman1937use,friedman1940comparison} and Holm's step down 
procedure can be used to test for significant differences between SparseStep 
and other methods \citep{holm1979simple}.

\begin{figure*}[tb]
	\def\FigureWidth{.98\textwidth}
	\centering
	\subfigure[MSE of $\bfs{\hat{\beta}}$]{%
		\includegraphics[width=\FigureWidth]{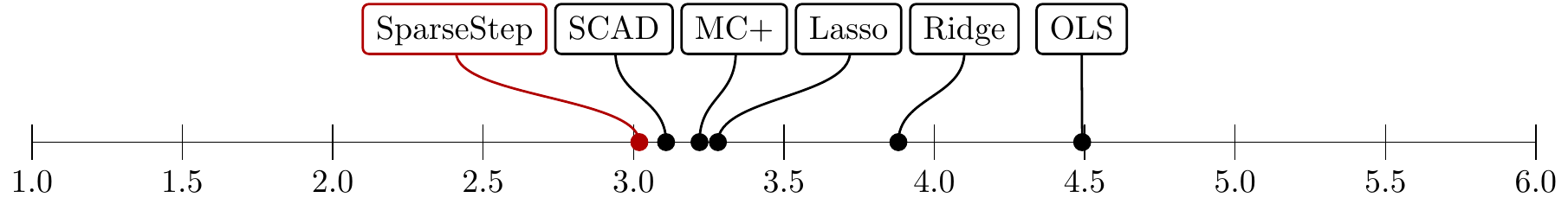}
		\label{fig:res_beta_mse_mse}
	}\qquad
	\subfigure[Sparsity hitrate $\bfs{\hat{\beta}}$]{%
		\includegraphics[width=\FigureWidth]{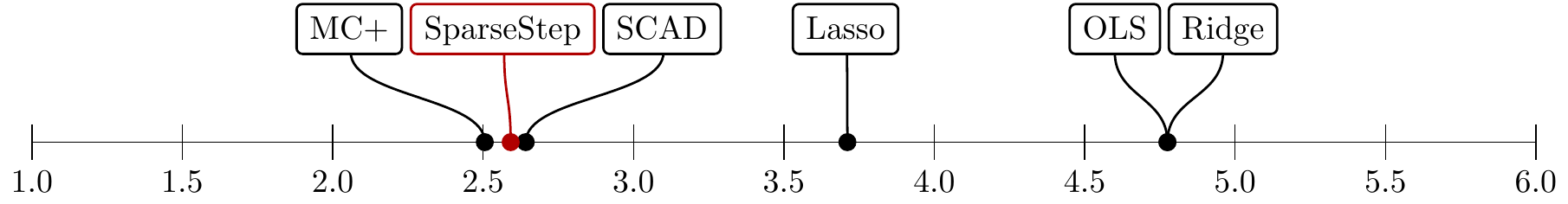}
		\label{fig:res_beta_mse_hit}
	}\qquad
	\subfigure[MSE of $\bft{\hat{y}}$ on test datasets]{%
		\includegraphics[width=\FigureWidth]{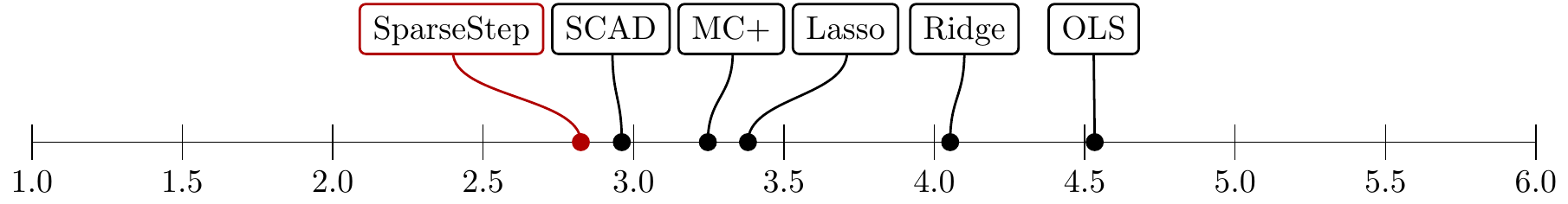}
		\label{fig:res_ytest_mse_mse}
	}\qquad
	\subfigure[Average computation time]{%
		\includegraphics[width=\FigureWidth]{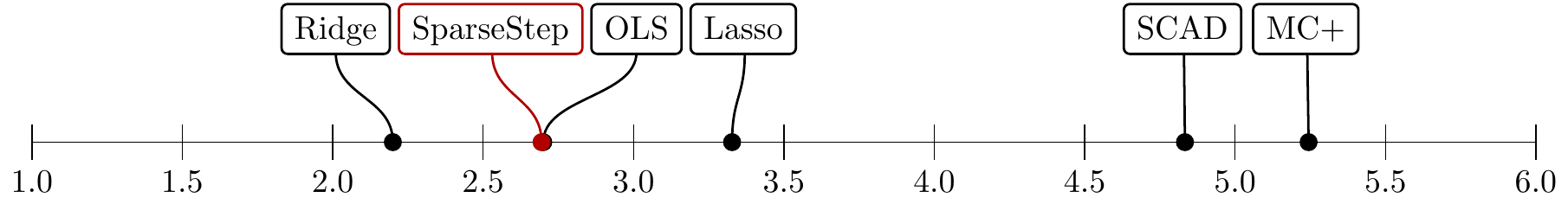}
		\label{fig:res_time}
	}
	\caption{Rank plots showing the average performance of the six 
		evaluated methods on the 180 simulated datasets. Points on the 
		line show the fractional rank of the methods averaged over all  
		datasets, with a smaller rank indicating a better performance.  
		It can therefore be seen that SparseStep performs very 
		favorably in both predicting the size and sparsity of 
		$\bfs{\beta}$, as well as on predicting the outcome 
		out-of-sample. Graph (d) shows the rank graph for the average 
		computation time for each method. \label{fig:res_ranks}}
\end{figure*}

Figure \ref{fig:res_ranks} shows the average ranks of the six evaluated 
methods on four different metrics. From Figure \ref{fig:res_beta_mse_mse}, 
which shows the average ranks on the MSE of $\bfs{\hat{\beta}}$, it can be 
seen that SparseStep is most often the best method for fitting $\bfs{\beta}$, 
followed closely by SCAD, MC+, and the Lasso. The sparsity hitrate of 
$\bfs{\hat{\beta}}$ is on average the best for the MC+ penalty, followed by 
SparseStep and SCAD, as illustrated in Figure \ref{fig:res_beta_mse_hit}. For 
the out-of-sample performance on the test data, shown in 
\ref{fig:res_ytest_mse_mse}, a similar order of the methods can be observed as 
for the MSE on $\bfs{\hat{\beta}}$, although the difference between SCAD and 
MC+ is larger here. SparseStep again outperforms the other methods on this 
measure.

Computation time was also measured for each method on each dataset. The rank 
plot of the average computation time per dataset is given in Figure 
\ref{fig:res_time}. It can be seen that SparseStep performs well on average.  
The average computation time of SparseStep for a single value of $\lambda$ is 
comparable to computing a single OLS solution. An important caveat with 
regards to the computation times is that in order to have the same CV splits 
for each method with a certain $\lambda$, the path algorithms of the Lasso, 
SCAD, and MC+ penalty could not be used. The computation time of these methods 
is therefore overestimated.

Apart from the ranks averaged over all datasets, it is also interesting to 
look at how often a method is the best method on a dataset and how often it is 
the worst method. Looking at the MSE of $\bfs{\hat{\beta}}$, Ridge is most 
often the best method, followed closely by the penalized methods. As expected 
OLS is most often the worst method in this regard.  Next, SparseStep most 
often obtains the highest sparsity hitrate of $\bfs{\beta}$, on 30 out of 180 
datasets. MC+ is the best method on this metric on 26 datasets and SCAD on 11 
datasets.  Finally, when considering the MSE on the outcome of the test 
dataset all the penalized methods are the best method with similar frequency.  
An exception to this is OLS, which is the best method on only 7 datasets and 
the worst method on 52 of them. MC+ is the worst method on 32 datasets, 
whereas SparseStep is the worst method the smallest number of times, on only 6 
datasets.  Clearly, OLS and Ridge only perform well on datasets without 
sparsity in $\bfs{\beta}$.  For the other methods no clear relationship 
between the dataset characteristics and the performance of the method could be 
found.

As suggested by \citet{demvsar2006statistical} an $F$-test can be done on the 
average ranks to evaluate if significant differences exist between the 
different methods. This is the case for the four measures discussed above, all 
with p-value $< 0.0001$. Furthermore, Holm's procedure can be performed to 
uncover significant differences between the methods and a reference method, in 
this case SparseStep. From this it is found that on the performance metrics 
other than computation time, SparseStep significantly outperforms OLS and 
Ridge, but that the difference between SparseStep and the other penalized 
methods is not significant at the 5\% level. On the computation time metric 
SparseStep significantly outperforms SCAD and MC+ at the 5\% level, but the 
caveat mentioned above should be taken into account here. The lack of a 
significant difference between SparseStep and SCAD and MC+ on the other 
metrics can be due to either a lack of any theoretical difference, or an 
insufficient number of datasets in the simulation study.

\section{Discussion}
\label{sec:discussion}
This paper introduces the SparseStep algorithm which induces sparsity in the 
regression problem by iteratively improving an approximation of the $\ell_0$ 
norm. An iterative majorization algorithm has been derived which is 
straightforward to implement. The practical relevance of SparseStep is 
evaluated through a thorough simulation study on 180 datasets with varying 
characteristics. Results of the simulation study indicate that SparseStep 
often outperforms existing methods, both for identifying the parameter vector 
$\bfs{\beta}$ as for out-of-sample prediction of the outcome variable.  Future 
research will focus on furthering the understanding of the theoretical 
properties of the SparseStep algorithm, such as the criteria for convergence 
to a global optimum.

\section*{Acknowledgements}
The computational tests of this research were performed on the Dutch National 
LISA cluster, and supported by the Dutch National Science Foundation (NWO).  
The authors thank SURFsara (\url{www.surfsara.nl}) for the support in using 
the LISA cluster.

\bibliographystyle{plainnat}
\bibliography{sparsestep}


\end{document}